\documentclass[11pt,reqno]{amsart}

\usepackage{fancyhdr}
\usepackage{graphicx}
\usepackage{amsmath,amssymb,amsthm,amsfonts,amsxtra,physics}

\usepackage[hyperfootnotes=true,
pdffitwindow=true,
plainpages=false,
pdfpagelabels=true,
pdfpagemode=UseOutlines,
pdfpagelayout=SinglePage,
hyperindex,]{hyperref}

\theoremstyle{plain}

\theoremstyle{definition}

\theoremstyle{remark}

\title[Finite size effects in real symmetric Wigner matrices]{Comment on "Finite size effects in the averaged eigenvalue density of Wigner random-sign real symmetric matrices" by G.S. Dhesi and M. Ausloos}
\author{Peter J. Forrester and Allan K. Trinh}
\address{Department of Mathematics and Statistics, 
	ARC Centre of Excellence for Mathematical \& Statistical Frontiers,
	University of Melbourne, Victoria 3010, Australia}
\email{pjforr@unimelb.edu.au}
\email{a.trinh4@student.unimelb.edu.au}
\date{\today}	

\begin{document}
\maketitle
\begin{abstract}
	The recent paper ``Finite size effects in the averaged eigenvalue density of Wigner random-sign real symmetric matrices" by G.S. Dhesi and M. Ausloos [Phys. Rev. E \textbf{93} (2016), 062115] uses the replica method to compute the $1/N$ correction to the Wigner semi-circle law for the ensemble of real symmetric random matrices with $0$'s down the diagonal, and upper triangular entries independently chosen from the values $\pm w$ with equal probability. We point out that the results obtained are inconsistent with known results in the literature, as well as with known large $N$ series expansions for the trace of powers of these random matrices. An incorrect assumption relating to the role of the diagonal terms at order $1/N$ appears to be the cause for the inconsistency.
	Moreover, results already in the literature can be used to deduce the $1/N$ correction to the Wigner semi-circle law for real symmetric random matrices with entries drawn independently from distributions $\mathcal D_1$ (diagonal entries) and $\mathcal D_2$ (upper triangular entries) assumed to be even and have finite moments. Large $N$ expansions for the trace of the $2k$-th power ($k=1,2,3$) for these matrices can be computed and used as  checks.
\end{abstract}

\section{Introduction}
The class of Wigner random matrices refers to real symmetric, or complex Hermitian, $N\times N$ matrices with entries on the diagonal chosen independently from a zero mean distribution $\mathcal D_1$, and upper triangular entries chosen independently from a finite mean, finite variance (the latter equal to $w^2$ say) distribution $\mathcal D_2$. Now scale the matrices by multiplying by $1/\sqrt{N}$, and define the scaled eigenvalue density $\bar{\rho}(\lambda)$ --- which apriori is a function of $N$, $\mathcal D_1$ and $\mathcal D_2$ --- by the requirement that $N\int_a^b \bar{\rho}(\lambda) \, d\lambda$ is equal to the expected number of eigenvalues in the interval $[a,b]$. It is a celebrated result (see e.g. \cite[Theorem 18.3.2 and Remark 18.3.3]{PS11}) that for a suitable class of test functions $\phi(\lambda) $,
\begin{equation}\label{eq:2.1}
	\lim_{N\to\infty} \int_{-\infty}^\infty \phi(\lambda)\bar{\rho}(\lambda)\, d\lambda = \int_{-2w}^{2w} \phi(\lambda)\rho_{\rm W,0}(\lambda) \, \, d\lambda,
\end{equation}
where
\begin{equation}\label{eq:2.1a}
	\rho_{\rm W,0}(\lambda)=\frac{1}{2\pi w^2}(4w^2-\lambda^2)^{1/2}.
\end{equation}
Due to the shape of (\ref{eq:2.1a}) when plotted as a graph, and the fact that results of this sort was first obtained by Wigner \cite{Wi55,Wi58}, the result implied by (\ref{eq:2.1}) and (\ref{eq:2.1a}) is referred to as the Wigner semi-circle law. Note that no property of $\mathcal  D_1, \mathcal  D_2$ beyond those stated affects the functional form (\ref{eq:2.1a}), which thus exhibits a type of universality.

Wigner's original 1955 paper \cite{Wi55} considered $(2N+1)\times (2N+1)$ real symmetric matrices, $J_{2N+1}$ say, with diagonal entries all zero, upper triangular entires independently chosen from the values $\pm w$ with equal probability. A result equivalent to (\ref{eq:2.1}) was derived by using the family of test functions $\phi(\lambda)=\lambda^p$ ($p=2,4,...$) which corresponds to the moments of the spectral density. Wigner's second paper, published in 1958 \cite{Wi58}, considered a wider class of random real symmetric matrices: independence of entries, fixed variance and bounded moments were shown by the same technique to be sufficient conditions, again based on the method of moments. The removal of the necessity of finite moments, beyond the variance, came later upon the refinement of the analysis of the Stieltjes transform of the spectral density, as introduced by Marchenko and Pastur \cite{MP67}.

In a recent work published in this journal by Dhesi and Ausloos \cite{DA16}, the spectral density of Wigner's original class of random matrices $J_N $  --- now without the restriction to $N$ odd --- was reconsidered. In particular for $N$ large but finite the leading correction term to the RHS of (\ref{eq:2.1}) was sought. Earlier \cite{DJ90}, the first author of \cite{DA16} together with Jones had found this leading correction in the case of the Gaussian orthogonal ensemble (GOE) --- real symmetric matrices with diagonal entries independent normal distributions $\text{N}[0,\sqrt{2}w]$ and off-diagonal entries independent normal distributions $\text{N}[0,w]$; see e.g.~\cite[\S 1.1]{Fo10}. Thus it was shown that for large $N$
\begin{equation}\label{eq:4.1}
	\int_{-\infty}^\infty \phi(\lambda)\bar{\rho}(\lambda)\, d\lambda 
	= \int_{-2w}^{2w} \phi(\lambda)\rho_{\rm W,0}(\lambda) \, d\lambda 
	+ \frac{1}{N} \int_{-\infty}^\infty \phi(\lambda)\rho_{\rm W,1}(\lambda) \, d\lambda
	+ \cdots
\end{equation}
where, with  $\chi_A=1$ for $A$ true, $\chi_A=0$ otherwise
\begin{equation}\label{eq:5.1}
	\rho_{\rm W,1}(\lambda) = \frac{1}{4}(\delta(\lambda+2w)+\delta(\lambda-2w))
	-\frac{1}{2\pi}\frac{\chi_{\abs{\lambda}<2w}}{(4w^2-\lambda^2)^{1/2}}
\end{equation}
(see also \cite{VZ84}, \cite{Jo98}, \cite{WF14}). We remark that in \cite{DA16} the symbol $J$ is used for our $w$.

The replica method from statistical physics was applied by Dhesi and Ausloos to calculate that in the case of the random matrices $J_N $, the asymptotic formula (\ref{eq:4.1}) should be modified to read \cite[eqns (52), (53), (54)]{DA16}
\begin{equation}\label{eq:5.2a}
	\int_{-\infty}^\infty \phi(\lambda)\bar{\rho}(\lambda)\, d\lambda 
	= \int_{-2w}^{2w} \phi(\lambda)\rho_{\rm W,0}(\lambda) \, d\lambda 
	+ \frac{1}{N} \int_{-\infty}^\infty \phi(\lambda)
	\left( \rho_{\rm W,1}^{(Q)}(\lambda)+\rho_{\rm W,1}^{(R)}(\lambda)
	\right)  d\lambda
	+ \cdots
\end{equation}
where $\rho_{\rm W,1}^{(Q)}(\lambda)$ is given by (\ref{eq:5.1}), while
\begin{equation}\label{eq:5.2b}
	\rho_{\rm W,1}^{(R)}(\lambda)=\frac{3}{8\pi} {(4w^2-\lambda^2)^{1/2} \over w^4}
	\left( (3\lambda^2-2w^2) - {2\lambda^2(\lambda^2-2w^2) \over 4w^2-\lambda^2 }
	\right) \chi_{\abs{\lambda}<2w}.
\end{equation}
With $m_p:=\int_{-\infty}^\infty \lambda^p\bar{\rho}(\lambda)\, d\lambda$ denoting the moments of the scaled eigenvalue density, define the moment generating function (Green's function) by
\begin{equation}\label{eq:5.2c}
	G(x)=\frac{1}{x} \sum_{p=0}^\infty \frac{m_p}{x^p} = \int_{-\infty}^\infty {\bar{\rho}(\lambda)\over x-\lambda}\, d\lambda.
\end{equation}
As a corollary of (\ref{eq:5.2a}), (\ref{eq:5.2b}) it is deduced that \cite[eqns (52), (53), (54)]{DA16}
\begin{equation}\label{eq:5.2d}
	G(x)=G_{\rm W,0}(x)+\frac{1}{N} \left(
	G_{\rm W,1}^{(Q)}(x)+G_{\rm W,1}^{(R)}(x)
	\right) + \cdots
\end{equation}
where
\begin{align}
	\label{eq:5.2e}
	G_{\rm W,0}(x)&={1\over 2w^2} \left( x-x(1-4w^2/x^2)^{1/2} \right),
	\\
	\label{eq:5.3a}
	G_{\rm W,1}^{(Q)}(x)&={(wG_{\rm W,0}(x))^3\over w(1-(wG_{\rm W,0}(x))^2)^2},
	\\
	\label{eq:5.3b}
	G_{\rm W,1}^{(R)}(x)&=-3{(wG_{\rm W,0}(x))^5\over w(1-(wG_{\rm W,0}(x))^2)}.
\end{align}
In our notation, we have factored the $1/N$ as seen in (\ref{eq:5.2d}), whereas in \cite{DA16} the $1/N$ factor is included in (\ref{eq:5.3a}), (\ref{eq:5.3b}). It is commented \cite[below (57)]{DA16} that the first of the terms in (\ref{eq:5.2d}) proportional to $1/N$ corresponds to the $1/N$ correction to $G(x)$ for the Gaussian orthogonal ensemble. It is similarly true that $\rho_{\rm W,1}^{(Q)}(\lambda)$ in \eqref{eq:5.2a} is the $1/N$ correction in the expansion of $\int_{-\infty}^\infty \phi(\lambda)\bar{\rho}(\lambda)\, d\lambda$ applied to the GOE \cite[below (53)]{DA16}.

In fact neither of the corrections (\ref{eq:5.2b}) nor (\ref{eq:5.3b}) to the GOE result are correct, and they are not consistent with each other. With regards to (\ref{eq:5.2b}), use of computer algebra gives
\begin{equation}\label{eq:R1}
	\int_{-2w}^{2w}\rho_{\rm W,1}^{(R)}(\lambda)\, d\lambda=-{3\over 4}.
\end{equation}
But we also have
\begin{equation}\label{eq:R2}
	\int_{-\infty}^\infty \bar{\rho}(\lambda)\, d\lambda = \int_{-\infty}^\infty \rho_{\rm W,0}(\lambda)\, d\lambda =1, \quad
	\int_{-\infty}^\infty \rho_{\rm W,1}^{(Q)}(\lambda)\, d\lambda=0.
\end{equation}
Substituting (\ref{eq:R1}), (\ref{eq:R2}) in (\ref{eq:5.2a}) with $\phi(\lambda)=1$ leads to a contradiction: the results (\ref{eq:R2}) imply that we must have
\begin{equation}\label{eq:R3}
	\int_{-2w}^{2w}\rho_{\rm W,1}^{(R)}(\lambda)\, d\lambda=0.
\end{equation}
On the other hand, expanding (\ref{eq:5.3b}) for large $x$, taking into consideration (\ref{eq:5.2e}) gives
\begin{equation}\label{eq:R4}
	G_{\rm W,1}^{(R)}(x) \sim -{3w^4\over x^5}
\end{equation}
implying upon recalling (\ref{eq:5.2c})  that (\ref{eq:R3}) holds, in contradiction to (\ref{eq:R1}).

Similar inconsistencies between (\ref{eq:5.2b}) and (\ref{eq:5.3b}) are present when considering the second moment.
From (\ref{eq:5.2b}), use of computer algebra
gives
\begin{equation}\label{eq:R3a}
	\int_{-2w}^{2w} \lambda^2 \rho_{\rm W,1}^{(R)}(\lambda)\, d\lambda= - 3w^2.
\end{equation}	
However (\ref{eq:R4}) together with (\ref{eq:5.2c}) imply
\begin{equation}\label{eq:R3b}
	\int_{-2w}^{2w} \lambda^2 \rho_{\rm W,1}^{(R)}(\lambda)\, d\lambda= 0.
\end{equation}	
We claim that the correct value is
\begin{equation}\label{Sa}
	\int_{-2w}^{2w} \lambda^2\rho_{\rm W,1}^{(R)}(\lambda)\, d\lambda=-2w^2.
\end{equation}
To see this, first note that
according to (\ref{eq:5.3a}) and (\ref{eq:5.2e}), for large $x$ we have $G_{\rm W,1}^{(Q)}(x)\sim w^2/x^3$ and thus
\begin{equation}\label{eq:R6}
	\int_{-\infty}^\infty \lambda^2\rho_{\rm W,1}^{(Q)}(\lambda)\, d\lambda=w^2.
\end{equation} 
Alternatively, this can be derived directly from (\ref{eq:5.1}).
Adding (\ref{Sa}) and (\ref{eq:R6}) we see from (\ref{eq:5.2a}) that (\ref{Sa}) implies the $1/N$ correction to the second moment is $-w^2$. To
check this, from the definition of the random matrices $\{ J_N\}$ --- symmetric matrices with diagonal entries equal to $0$ and upper triangular elements chosen from the values $\pm w$ with equal probability --- we have that
\begin{equation}\label{eq:R7}
	\langle\Tr J_N^2\rangle = N(N-1)w^2,
\end{equation}
which indeed displays the claimed $1/N$ correction.

Results already in the literature \cite{BY05, KKP96} allow us to correct both (\ref{eq:5.2b}) and (\ref{eq:5.3b}),
\begin{align}
	\label{eq:R9}
	\rho_{\rm W,1}^{(R)\ast}(\lambda)
	&={1\over 2\pi w}\left(
	-2T_2(\lambda/2w)-2T_4(\lambda/2w)
	\right)
	{\chi_{\abs{\lambda}<2w}\over \sqrt{1-(\lambda/2w)^2}}
	\\
	\label{eq:R10}
	G_{\rm W,1}^{(R)\ast}(x)
	&= -{2(wG_{\rm W,0}(x))^3\over w(1-(wG_{\rm W,0}(x))^2) } - {2(wG_{\rm W,0}(x))^5\over w(1-(wG_{\rm W,0}(x))^2) }
\end{align}
where the $\ast$ on the superscript of the notation is to distinguish these corrected formulas from the incorrect ones. In (\ref{eq:R9}), $T_2$ and $T_4$ denote the second and fourth order Chebychev polynomials,
\begin{equation*}
	T_2(x) = 2x^2-1, \quad T_4(x)=8x^4-8x^2+1.
\end{equation*}
We will subject (\ref{eq:R9}) and (\ref{eq:R10}) to a number of checks in the next section.

The real symmetric matrices $J_N$ considered in \cite{DA16} may be extended to more general to real symmetric Wigner matrices
$X = [x_{jk} ]$ by allowing for non-zero entries on the diagonal. We suppose each entry is chosen independently from a distribution $\mathcal D_1$ with zero mean and standard deviation $v^2$. The distribution $\mathcal D_2$ of the off-diagonal entries --- which for $\{J_N\}$ has the $2k$-th moment equal to $w^{2k}$ for each $k=1,2,...$ --- may also be generalised by allowing for a general value of the fourth cumulant
\begin{equation*}
	\kappa_4 = \langle x_{j,k}^4 \rangle -3\langle x_{j,k}^2 \rangle^2 = \langle x_{j,k}^4 \rangle -3w^4.
\end{equation*}
In the generalised setting (assuming too that both $\mathcal D_1$ and $\mathcal D_2$ are even with all moments finite), the existing literature \cite{Gi89,KKP96,BY05} tells us that the expansion (\ref{eq:4.1}) depends on $v^2$, $w^2$ and $\kappa_4$, but not on higher moments of 
$\mathcal D_1$ and $\mathcal D_2$. The expansion (\ref{eq:5.2a}) again applies with $\rho_{\rm W,1}^{(Q)}(\lambda)$ given by (\ref{eq:5.1}), but $\rho_{\rm W,1}(\lambda)$ replaced by the quantity
\begin{equation}\label{eq:S1}
	\rho_{\rm W,1}^{(R)\#}(\lambda)
	={1\over 2\pi w}\left(
	\left( \frac{v^2}{w^2}-2 \right) T_2(\lambda/2w) + \frac{\kappa_4}{w^4}T_4(\lambda/2w)
	\right)
	{\chi_{\abs{\lambda}<2w}\over \sqrt{1-(\lambda/2w)^2}}.
\end{equation}
And for the expansion (\ref{eq:5.2d}), $G_{\rm W,1}^{(Q)}(x)$ is again given by (\ref{eq:5.3a}) while $G_{\rm W,1}^{(R)}(x)$ is to be replaced by the quantity
\begin{equation}\label{eq:S2}
	G_{\rm W,1}^{(R)\#}(x)
	= {(v^2/w^2-2)(wG_{\rm W,0}(x))^3\over w(1-(wG_{\rm W,0}(x))^2) } + {(\kappa_4/w^4)(wG_{\rm W,0}(x))^5\over w(1-(wG_{\rm W,0}(x))^2) }.
\end{equation}
The fact that (\ref{eq:S1}) and (\ref{eq:S2}) depend on the variance $v^2$ of the diagonal elements tell us that a key claim in \cite{DA16}, asserting that at order $1/N$ there is no such dependence, is incorrect. Thus the first sentence of the paragraph including (1) in \cite{DA16} reads: ``The present paper is still devoted to the calculation of the averaged eigenvalue density..., with vanishing diagonal elements --- though it will be shown that this constraint is rather irrelevant to order $1/N$." This claim is (essentially) repeated as the final sentence of the Appendix, the latter in turn being added in response to a comment of a referee; see the sentence after  \cite[Eq.~(13)]{DA16}.

\section{Comparison with exact moments}
Cicuta \cite{Ci05} has used graphical methods to compute the exact form of $\langle {\rm Tr} \, S_0^p\rangle$ ($p=1,2,...,8$) in the case that the $N\times N$ random real symmetric matrix $S_0$ has entries equal to zero on the diagonal, and identically distributed elements above the diagonal with moments $\langle w^p\rangle$ ($p=1,2,...$). The special case $\langle w^{2k+1}\rangle=0$, $\langle w^{2k}\rangle =w^{2k}$ ($k=1,2,...$) corresponds to the random matrix $J_N$ considered in \cite{DA16}. In the case that the odd moments of the entries above the diagonal are zero, and the even powers arbitrary we read off from \cite{Ci05} that
\begin{align}
	\langle \Tr J_N^2 \rangle 
	&= N(N-1)\langle w^2\rangle \nonumber
	\\
	\langle \Tr J_N^4 \rangle 
	&= N(N-1)\langle w^4\rangle + 2N(N-1)(N-2)\langle w^2\rangle^2 \nonumber
	\\
	\langle \Tr J_N^6 \rangle 
	&= N(N-1)\langle w^6\rangle + 6N(N-1)(N-2)\langle w^4\rangle \langle w^2\rangle \nonumber
	+ N(N-1)(N-2)(5N-11)\langle w^2\rangle^3
	\\
	\langle \Tr J_N^8 \rangle 
	&= N(N-1)\langle w^8\rangle + N(N-1)(N-2)(8\langle w^6\rangle\langle w^2\rangle + 6\langle w^4\rangle^2) \nonumber
	\\
	&\quad + 28N(N-1)(N-2)^2\langle w^4\rangle\langle w^2\rangle^2 + N(N-1)(N-2)(N-3)(14N-19)\langle w^2\rangle^4.
	\label{eq:7.1}
\end{align}
Restricting to the first two leading orders in $N$ and recalling that for $J_N$, $\langle w^{2k}\rangle =w^{2k}$, these reduce to
\begin{align*}
	\langle \Tr J_N^2 \rangle 
	&= N^2 w^2 - N w^2
	\\
	\langle \Tr J_N^4 \rangle 
	&= 2N^3 w^4-5N^2w^4+\cdots
	\\
	\langle \Tr J_N^6 \rangle 
	&= 5N^4w^6-20N^3w^6+\cdots
	\\
	\langle \Tr J_N^8 \rangle 
	&= 14N^5w^8-75N^4w^8+\cdots.
\end{align*}
The coefficients of the subleading term are in keeping with the moments
\begin{equation*}
	\int_{-\infty}^\infty \lambda^{2k} \left( \rho_{\rm W,1}^{(Q)}(\lambda)+\rho_{\rm W,1}^{(R)\ast}(\lambda)
	\right) \, d\lambda
	=
	\begin{cases}
	-w^2, \quad &k=1 \\
	-5w^4, \quad &k=2 \\
	-20w^6, \quad &k=3 \\
	-75w^8, \quad &k=4,
	\end{cases}
\end{equation*}
as implied by the functional forms (\ref{eq:5.1}) and (\ref{eq:R9}). They are similarly consistent with the coefficients in the large $x$ expansion
\begin{equation*}
	G_{\rm W,1}^{(Q)}(x)+G_{\rm W,1}^{(R)\ast}(x) = -\frac{w^2}{x^3}-\frac{5w^4}{x^5}-\frac{20w^6}{x^7}-\frac{75w^8}{x^9}-\cdots
\end{equation*}
as implied by (\ref{eq:5.3a}), (\ref{eq:5.2e}) and (\ref{eq:R10}).

Generalisation of (\ref{eq:7.1}) to the case that the diagonal entries are chosen from the same distribution as the off-diagonal entries was given in \cite{Ci05}. These results, at least up to the trace of the sixth power, can themselves be generalised so that the diagonal entries are sampled from a distinct distribution with moments $\langle v^p\rangle$ ($p=1,2,...$). Choosing too all odd moments to vanish we find
\begin{align*}
	\langle \Tr S^2 \rangle 
	&= N^2\langle w^2\rangle + N(\langle v^2\rangle-\langle w^2\rangle)
	\\
	\langle \Tr S^4 \rangle 
	&= 2N^3\langle w^2\rangle^2 
	+ N^2(4\langle w^2\rangle\langle v^2\rangle+\langle w^4\rangle-6\langle w^2\rangle^2)
	+ N(\langle v^4\rangle-\langle w^4\rangle-4\langle w^2\rangle\langle v^2\rangle+4\langle w^2\rangle^2)
	\\
	\langle \Tr S^6 \rangle 
	&= 5N^4\langle w^2\rangle^3 + N^3(15\langle w^2\rangle^2\langle v^2\rangle+6\langle w^2\rangle\langle w^4\rangle-26\langle w^2\rangle^3)
	\\
	& \quad + N^2(6\langle w^2\rangle\langle v^4\rangle + 3\langle w^2\rangle\langle v^2\rangle^2 +9\langle w^4\rangle\langle v^2\rangle -45\langle w^2\rangle^2\langle v^2\rangle + \langle w^6\rangle - 18\langle w^2\rangle\langle w^4\rangle + 43\langle w^2\rangle^3)
	\\
	& \quad + N(\langle v^6\rangle - 6\langle w^2\rangle\langle v^4\rangle - 3\langle w^2\rangle\langle v^2\rangle^2 - 9\langle w^4\rangle\langle v^2\rangle +30\langle w^2\rangle^2\langle v^2\rangle - \langle w^6\rangle + 12\langle w^2\rangle\langle w^4\rangle - 22\langle w^2\rangle^3).
\end{align*}
With $\langle w^2\rangle=w^2$, $\langle v^2\rangle=v^2$ and $\kappa_4=\langle w^4\rangle-3w^4$, restricting to the first two leading orders these read
\begin{align}
\langle \Tr S^2 \rangle 
&= N^2w^2+N(v^2-w^2) \nonumber
\\
\langle \Tr S^4 \rangle 
&= 2N^3w^4+N^2(4w^2v^2+\kappa_4-3w^4)+\cdots \nonumber
\\
\langle \Tr S^6 \rangle 
&= 5N^4w^6 +N^3(15w^4v^2+6w^2\kappa_4-8w^6)+\cdots.
\label{eq:9.1}
\end{align}
The coefficients of the subleading terms in (\ref{eq:9.1}) are consistent with the moments
\begin{align*}
	\int_{-\infty}^\infty \lambda^{2k}
	\left(
	\rho_{\rm W,1}^{(Q)}(\lambda) - { T_2(\lambda/2w)\chi_{\abs{\lambda}<2w}\over \pi w\sqrt{1-(\lambda/2w)^2} }
	\right)\, d\lambda
	&=
	\begin{cases}
	-w^2, \quad &k=1 \\
	-3w^4, \quad &k=2 \\
	-8w^6, \quad &k=3
	\end{cases}
	\\
	{v^2\over 2\pi w^3}\int_{-2w}^{2w} \lambda^{2k} { T_2(\lambda/2w)\over \sqrt{1-(\lambda/2w)^2} } \, d\lambda
	&=
	\begin{cases}
	v^2, \quad &k=1 \\
	4w^2 v^2, \quad &k=2 \\
	15w^4 v^2, \quad &k=3
	\end{cases}
	\\
	{\kappa_4 \over 2\pi w^5}\int_{-2w}^{2w} \lambda^{2k} { T_4(\lambda/2w)\over \sqrt{1-(\lambda/2w)^2} } \, d\lambda
	&=
	\begin{cases}
	0, \quad &k=1 \\
	\kappa_4, \quad &k=2 \\
	6w^2\kappa_4, \quad &k=3.
	\end{cases}
\end{align*}
It is also the case that for large $x$
\begin{align*}
	{(wG_{\rm W,0}(x))^3 \over w(1-(wG_{\rm W,0}(x))^2)^2 } - { 2(wG_{\rm W,0}(x))^3 \over w(1-(wG_{\rm W,0}(x))^2) }
	&=
	-\frac{w^2}{x^3} - \frac{3w^4}{x^5} - \frac{8w^6}{x^7} - \cdots
	\\
	{ v^2(wG_{\rm W,0}(x))^3 \over w^3(1-(wG_{\rm W,0}(x))^2) }
	&=
	\frac{v^2}{x^3} + \frac{4w^2x^2}{x^5} + \frac{15w^4v^2}{x^7} + \cdots
	\\
	{\kappa_4 (wG_{\rm W,0}(x))^5 \over w^5(1-(wG_{\rm W,0}(x))^2) }
	&= \frac{\kappa_4}{x^5} + \frac{6w^2\kappa_4}{x^7} + \cdots
\end{align*}
which are similarly consistent with (\ref{eq:5.2d}), upon replacing $G_{\rm W,1}^{(R)}$  by $G_{\rm W,1}^{(R)\#}$, and (\ref{eq:9.1}).

\section*{Acknowledgements}
This work is part of a research program supported by the Australian Research Council (ARC) through the ARC Centre of Excellence for Mathematical and Statistical frontiers (ACEMS). PJF also acknowledges partial support from ARC grant DP170102028, and AKT acknowledges the support of a Melbourne postgraduate award.

\providecommand{\bysame}{\leavevmode\hbox to3em{\hrulefill}\thinspace}
\providecommand{\MR}{\relax\ifhmode\unskip\space\fi MR }
\providecommand{\MRhref}[2]{%
  \href{http://www.ams.org/mathscinet-getitem?mr=#1}{#2}
}
\providecommand{\href}[2]{#2}

\end{document}